\documentclass[aps,
    amsmath,
    amssymb,
    prl,
    twocolumn,
    superscriptaddress,
    floatfix,
    nofootinbib,
    nobibnotes]{revtex4-2}

\usepackage{graphicx}
\usepackage{bm}
\usepackage{xcolor}
\usepackage[colorlinks=true,citecolor=blue,linkcolor=blue,urlcolor=blue]{hyperref}
\usepackage{orcidlink}
\usepackage[export]{adjustbox}

\graphicspath{{figures/}}

\usepackage{booktabs}

\newcommand{\PL}{P_{\mathrm{L}}}

\newcommand{\Ord}{\mathcal{O}}
\newcommand{\astar}{\alpha^{*}}

\begin{document}

\title{Correlated decoding loses the surface code's effective distance below threshold}

\author{C\'esar Hernando-Fuente\orcidlink{0009-0007-9438-9908}}
\altaffiliation[Present address: ]{IQM Quantum Computers, Munich, Germany}
\email[E-mail: ]{cesar.hernando22@gmail.com}
\affiliation{$\langle aQa^{L}\rangle$ Applied Quantum Algorithms and Instituut-Lorentz, Leiden University, Niels Bohrweg 1, 2333 CA Leiden, The Netherlands}

\author{Joonas Majaniemi\orcidlink{0000-0003-3389-8294}}
\affiliation{Riverlane, 59 St Andrews Street, CB2 3BZ Cambridge, United Kingdom}

\author{Evert van Nieuwenburg\orcidlink{0000-0003-0323-0031}}
\affiliation{$\langle aQa^{L}\rangle$ Applied Quantum Algorithms and Instituut-Lorentz, Leiden University, Niels Bohrweg 1, 2333 CA Leiden, The Netherlands}

\author{Felix Frohnert\orcidlink{0000-0003-3717-6352}}
\affiliation{$\langle aQa^{L}\rangle$ Applied Quantum Algorithms and Instituut-Lorentz, Leiden University, Niels Bohrweg 1, 2333 CA Leiden, The Netherlands}

\date{\today}

\begin{abstract}
Correlated decoding improves matching and clustering surface-code decoders by reweighting one decoding subgraph conditioned on the other's correction. 
We show that standard reweighting can cause correlated decoders to fail on size-$(d-1)/2$ fault configurations that a distance-$d$ code should correct, reducing error suppression deep below threshold. 
Under circuit-level noise, a single first-pass edge can discount several correlated edges, even though their joint occurrence requires additional independent faults, so the decoder favors an over-discounted correction.
Regularizing the reweighting rule restores the nominal distance scaling and a clear correlated-decoding advantage.
\end{abstract}

\maketitle

\textit{Introduction.}---
Fault-tolerant quantum computation (FTQC) relies on classical decoding of stabilizer measurement outcomes to infer and correct errors~\cite{terhal2015quantum,dennis2002topological}.
In schemes based on repeated syndrome extraction, this decoding must keep pace with the cycle time of the quantum processor to avoid a growing backlog of unprocessed syndrome data~\cite{terhal2015quantum,battistel2023real, skoric2023parallel,barber2025real} while maintaining high accuracy~\cite{google2023suppressing,google2025quantum}.
Decoders based on tree search~\cite{beni2025tesseract} and machine learning~\cite{bausch2024learning,senior2025scalable} lead accuracy benchmarks, but efficient graph-based decoders such as minimum-weight perfect matching (MWPM)~\cite{higgott2022pymatching,higgott2025sparse} and Union-Find (UF)~\cite{delfosse2021almost,huang2020fault} remain the leading candidates for real-time decoding, thanks to fast and scalable hardware implementations~\cite{liyanage2023scalable, liynage2024helios, ziad2025local}.
Although recent experimental results have demonstrated real-time decoding of logical qubits in small-scale settings~\cite{google2023suppressing, google2025quantum, bluvstein2024logical}, reaching the logical error rates required for FTQC demands substantial improvements in qubit hardware as well as highly scalable, efficient and accurate classical decoders.

Surface codes~\cite{bravyi1998quantum, kitaev2003fault, fowler2012surface} are a primary candidate for FTQC because of their local stabilizer structure and the availability of efficient syndrome decoders. 
Graph-based decoders such as MWPM and UF decompose the surface code syndromes into $X$- and $Z$-type weighted subgraphs corresponding to the $X$- and $Z$-stabilizer measurements, allowing the two components to be decoded independently. 
However, the decomposition procedure discards essential information about error mechanisms that violate both stabilizer types, such as single-qubit $Y$-errors and cross-basis two-qubit gate errors, reducing decoding accuracy~\cite{fowler2013optimal}.
To address this loss of information, correlated protocols proposed in Refs.~\cite{fowler2013optimal,Paler_2023} use multiple decoding passes to achieve lower logical error probabilities than a baseline decoder across a range of physical error rates while only increasing the decoding latency by a constant factor.

Central to the correlated strategies is a graph reweighting policy that, after an initial decoding pass, updates the prior weights $w_\mu=\log[(1-p_\mu)/p_\mu]$ of each subgraph conditioned on the selected edges in the opposite subgraph. 
Here $p_\mu$ is the prior probability assigned to graph edge $e_\mu$; a single error mechanism can induce edges in both subgraphs.
A second decoding pass uses the updated weights to account for possible correlated error mechanisms. 
The two-pass protocol is decoder-agnostic, requiring only a decomposition of hyperedge error mechanisms into edges and the ability to reweight error priors conditioned on the decoder output.
In this work, we study Correlated Matching (CM), the two-pass variant of Sparse Blossom~\cite{higgott2025sparse}, an efficient implementation of MWPM for quantum error correction workloads~\cite{higgott2022pymatching,edmonds1965paths}, and the Local Clustering Decoder (LCD)~\cite{ziad2025local}, an adaptive weighted UF decoder optimized for efficient on-the-fly decoding graph updates. 

Surface code decoding benchmarks have largely focused on physical error rates accessible to current superconducting-qubit hardware, $p\sim10^{-3}$ to $10^{-2}$~\cite{google2023suppressing, google2025quantum,bausch2024learning,senior2025scalable}. 
Correlated decoding has been assessed mainly within this window, where its accuracy advantage over independent $X$-$Z$ decoding has been demonstrated. 
However, to avoid prohibitive encoding overheads~\cite{gidney2023yokedsurfacecodes}, FTQC demands operation deep below threshold, where the effectiveness of correlated decoding techniques has not yet been extensively studied.
Addressing this question is crucial not only as hardware improves but also for hierarchical architectures, in which an outer code layer may operate far below its threshold~\cite{pattison2023hierarchical,bluvstein2024logical}.
The performance of correlated decoders in the deep-subthreshold regime of $p<10^{-3}$ is therefore highly relevant for the long-term feasibility of FTQC systems.

\begin{figure*}[!t]
    \centering
    \includegraphics[width=\linewidth]{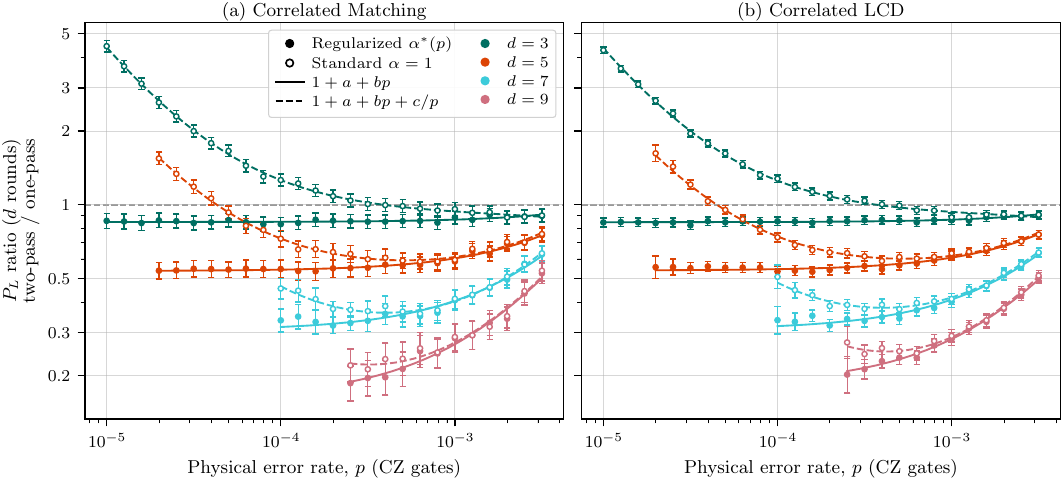}
    \caption{\textbf{Breakdown and regularization of correlated decoding below threshold.}
    Ratio of the $d$-round logical error probability of two-pass correlated decoding to that of the corresponding one-pass decoder under SI1000 noise.
    \textbf{(a)} Correlated matching (CM) and regularized correlated matching (RCM), normalized by uncorrelated matching, for $d=3,5,7,9$.
    \textbf{(b)} Correlated and regularized LCD, normalized by uncorrelated LCD.
    Open markers denote the standard update, $\alpha=1$, and filled markers the regularized update with $\astar(p)$.
    Dashed curves show fits containing the low-$p$ term $c/p$, whereas solid curves show the regular form $1+a+bp$.
    The horizontal line marks equal performance; values below unity favor the correlated decoder.
    Error bars are calculated as described in Appendix~\ref{sm:uncertainty}.
    The regularization $\astar$ restores a correlated decoding advantage across the error rate range studied.}
    \label{fig:degradation}
\end{figure*}

In this work, we show that standard correlated decoding loses its near-threshold advantage over uncorrelated decoding at low $p$: both correlated matching and correlated LCD can perform worse than their uncorrelated variants in the deep-subthreshold error regime.
Under circuit-level noise, the standard correlation reweighting rule can discount several edges conditioned on the same first-pass edge, leading to a second-pass correction whose assigned weight neglects the cost of additional independent faults required to explain it. 
To address these failures, we introduce an error-rate-dependent regularization parameter that increases the cost of each reweighted edge, preventing the decoder from selecting too many such edges. 
This recovers the appropriate asymptotic distance scaling of the logical error probability of correlated decoders while retaining a constant accuracy advantage over the uncorrelated variants deep below threshold.
Moreover, the regularization process retains the high efficiency of the standard correlated protocols, requiring only a one-time calculation for the new weights at decoder initialization and leaving the core decoding subroutine untouched.
  
\textit{Setup.}---
To benchmark decoding performance, we simulate distance-$d$ rotated surface code quantum memory circuits with $r=d$ syndrome-extraction rounds, serving as a standard proxy for the logical fidelity achievable in more complex experiments such as lattice surgery and non-Clifford logic~\cite{litinski2019game,gidney2024magic}. 
Circuits are simulated under SI1000 noise~\cite{gidney2021fault}, a superconducting-inspired circuit-level noise model parameterized by the two-qubit depolarizing error probability $p$. 
The effects observed are not unique to the SI1000 model, as shown by supporting simulations under uniform depolarizing noise in Section~\ref{sm:uniform_dep_noise} of the Appendix. 
We use Stim~\cite{gidney2021stim} to generate the circuits, sample shots, and build detector error models (DEMs). 
DEMs for correlated matching are decomposed into graphlike components using $\textrm{X-Z}$ annotations following the implementation in the \textsc{Tesseract} repository~\cite{beni2025tesseract}.

In the following, we describe the correlated decoding pipeline shared by correlated matching, implemented in \textsc{PyMatching v2.3.0}~\cite{higgott2025sparse}, and correlated LCD, a decoder proprietary to Riverlane~\cite{ziad2025local}.
Let $\mathcal M^{(1)}=\mathcal M_X^{(1)}\cup\mathcal M_Z^{(1)}$ denote the set of edges selected in the first pass. 
For each selected edge $e_\mu\in\mathcal M^{(1)}$ in each of the $X$- and $Z$-subgraphs, the update treats each selected edge as having occurred and updates the probability of its correlated edges (sharing a hyperedge mechanism in the DEM) conditioned on this information.
The conditional probability $p'_\nu$ assigned to an edge $e_\nu$ correlated with a selected edge $e_\mu$, and its updated weight, are 
\begin{equation}
\begin{gathered}
    p'_\nu = P(e_\nu\mid e_\mu)
    = \frac{P(e_\nu \land e_\mu)}{P(e_\mu)}
    = \frac{p_{\nu\mu}}{p_\mu}, \\[6pt]
    w'_\nu = \log\frac{1-p'_\nu}{p'_\nu},
\end{gathered}
\label{eq:CM}
\end{equation}
where $p_{\nu\mu}=P(e_\nu \land e_\mu)$ is obtained from the non-decomposed DEM and $p_\mu$ is the marginal probability of edge $e_\mu$ occurring.
Further details of the reweighting protocol are explained in Section~\ref{sm:weight-scales} of the Appendix.

Under phenomenological noise, there is a one-to-one mapping from $e_\mu$ to $e_\nu$ (except on the code boundaries); however, in a circuit-level setting, $e_\mu$ can be shared by multiple independent error mechanisms, each with its own partner edge $e_\nu$.
It is in this latter setting that the update rule breaks down.

\textit{Breakdown below threshold.}---
Figure~\ref{fig:degradation} shows the ratio of logical error probabilities between the correlated and uncorrelated decoders in a $d$-round memory simulation.
A ratio above $1$ means that the correlated decoder fails more often than the uncorrelated decoder; a ratio below $1$ indicates an accuracy advantage. 
For correlated matching, Fig.~\ref{fig:degradation}(a), the near-threshold improvement increases with code distance.
Below $p\simeq5\times10^{-4}$, however, the gap between the correlated and uncorrelated decoders shrinks for every $d$ as $p\rightarrow0$.
Especially notable is the $d=3$ case, where the correlated decoder performs worse than an uncorrelated variant at $p<4\times10^{-4}$, and at $p=10^{-5}$ reaches an error rate $\sim4.4\times$ higher than the uncorrelated reference. 
At $d=5$, the correlated decoding ratio crosses unity near $p\simeq4\times10^{-5}$ and reaches roughly $1.5$ at $p\simeq2\times10^{-5}$.
The same deterioration appears for correlated LCD, with a near-identical increase in the correlated decoder failure rates compared to the uncorrelated decoders at low $p$, as shown in Fig.~\ref{fig:degradation}(b).

This relative degradation grows without bound as $p\rightarrow0$, where fault configurations containing the fewest independent error mechanisms dominate. 
At $p=10^{-5}$, over $80\%$ of the $d=3$ correlated decoder's failures stem from syndromes that the uncorrelated decoder decodes correctly.
At $d=7$ and $d=9$, the computational cost prevents us from resolving the crossover above unity at low $p$, but the accuracy degradation is already visible. 
The close agreement between the correlated matching and correlated LCD results, as well as the presence of the low-\(p\) upturn at higher distances, indicates that the failure originates in the reweighting protocol rather than
in either decoder.

\textit{Loss of effective code distance.}---
Figure~\ref{fig:lep_vs_p} of the Appendix shows the logical error probability $\PL$ vs $p$ for correlated MWPM and LCD, showing the probabilities underlying the ratios in Fig.~\ref{fig:degradation}.
As $p\rightarrow0$, correlated decoding departs from the nominal scaling $\PL\propto p^{(d+1)/2}$, which the uncorrelated decoder maintains. 
Instead of showing the expected error suppression, at low $p$ the correlated decoders trend towards $\PL\propto p^{(d-1)/2}$, indicating a unit loss of effective distance.
Table~\ref{tab:exponents} in the Appendix shows that error suppression exponents for the correlated decoders at $d=3$ and $d=5$ are closer to  $(d-1)/2$ than to the expected value $(d+1)/2$ when $p\sim10^{-5}$, indicating that the decoder fails on a significant number of size-$\frac{d-1}{2}$ fault configurations, which a distance-$d$ code is expected to correct.

To test whether size-$\frac{d-1}{2}$ fault configurations cause these failures, we compare their contributions to the logical error probability with those of size-$\frac{d+1}{2}$ configurations at $d\in\{3,5,7\}$ in Fig.~\ref{fig:low_size_p_L_contrib}. 
We find that the logical error probability contribution of size-$\frac{d-1}{2}$ faults deep below threshold is orders of magnitude higher for CM than for MWPM---confirming that size-$\frac{d-1}{2}$ errors are the source of the correlated decoder's loss of distance. 
Near threshold, higher-weight errors mask the lowest-size error contributions due to their higher multiplicity, resulting in a lower $\PL$ for CM than for MWPM in the high-$p$ regime.
However, as $p\rightarrow0$, $\PL$ is dominated by the lowest-size failure modes, explaining the degradation of the correlated decoding accuracy deep below threshold.

\textit{Origin of the correlated decoder failures.}---
Although the posterior reweighting strategy of Eq.~\eqref{eq:CM} works well under phenomenological noise, where any bulk $e_\mu$ is correlated with precisely one $e_\nu$, circuit-level noise gives rise to multiple error mechanisms that each share $e_\mu$ with different partner edges $e_\nu$.
At low $p$, a selected first-pass edge $e_\mu$ is likely to have arisen from only one of its associated error mechanisms. 
However, the protocol assigns each partner edge $e_\nu$ a weight of $\Ord(1)$, while prior weights scale as $\Ord(\log(1/p))$. This makes corrections containing several edges discounted by the same source edge artificially cheap. 
This, in turn, leads to longer correction strings for the correlated decoder than are physically sound and allows size-$\frac{d-1}{2}$ faults to cause logical failures.

The mechanism already appears at $d=3$, where a Pauli $XY$ error on a $CZ$ gate produces two correlated boundary edges, one in each subgraph. 
\begin{center}
  \includegraphics[width=\linewidth]{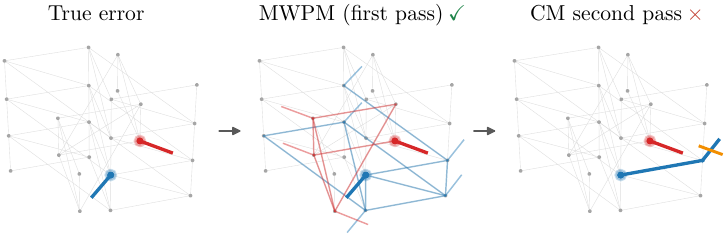}
\end{center}

The first pass recovers both correctly, but the $X$-type (red) edge discounts several $Z$-type (blue) edges at once. 
The second pass selects a chain of two discounted edges instead of the first-pass edge, whose updated weight exceeds their combined weight.  
One of the two second-pass $Z$-edges (marked with an orange line) anticommutes with the reference logical observable $Z_L$, producing a logical error.
A number of additional examples are shown in the Appendix, Section~\ref{sm:failure_examples}.

The problem stems from the decoder treating edges as independent mechanisms, adding the weights (i.e., multiplying probabilities) of the two $Z$-edges that have been discounted due to the same source $X$-edge.
This correction is valid within the graph error model, but it neglects the fact that, assuming the source edge occurred, only one of the two $Z$-edges can be attributed to the hyperedge mechanism involving the source edge.
If both $Z$-edges used in the second-pass correction are attributed to two independent hyperedge mechanisms, the first-pass source edge $e_{\mu}$ is no longer explained.
Therefore, one of the $Z$-edges used in the second-pass correction must be either a single edge mechanism or part of a hyperedge mechanism that does not involve the source edge.
Assuming that the $X$-edge $e_\mu$ occurs, we define the conditional probability of Eq.~\eqref{eq:CM} for a partner $Z$-edge $e_\nu$ as $q_\nu\equiv P(e_\nu\mid e_\mu)$.
The probability that a $Z$-edge $e_\sigma$ is produced by a mechanism not involving $e_\mu$ is given by $r_\sigma \equiv P(e_\sigma\land\lnot e_\mu)=P(e_\sigma)-P(e_\mu\land e_\sigma)=\Ord(p)$.
To leading order in $p$, the probability that both $Z$-edges $e_\nu$ and $e_\sigma$ occur is:
\begin{equation}
P(e_\nu\land e_\sigma\mid e_\mu)\simeq q_\nu r_\sigma + q_\sigma r_\nu = \Ord(p),
\label{eq:joint-true}
\end{equation}
whereas the second pass of MWPM (or LCD) implicitly assigns it
\begin{equation}
P(e_\nu\land e_\sigma\mid e_\mu) = q_\nu q_\sigma = \Ord(1).
\label{eq:joint-assumed}
\end{equation}
For the two-partner correction considered here, the decoder therefore underestimates the cost by
\begin{equation}
    \Delta w = \log\left(\frac{q_\nu q_\sigma}{q_\nu r_\sigma+q_\sigma r_\nu}\right)\simeq\log(1/p),
\end{equation}
the weight of one additional fault.
A size-$\frac{d-1}{2}$ fault configuration can then be explained by a logically nonequivalent correction that the decoder sees as no more costly than the correct one.
Table~\ref{tab:exponents} and Figure~\ref{fig:lep_vs_p} confirm the resulting unit loss of effective distance.
The resulting decision condition for the decoder is derived in Eq.~\eqref{eq:dw-dec} in the appendix.
This mechanism does not appear in the original study of Ref.~\cite{fowler2013optimal}, which uses a different circuit construction for the unrotated planar code. For the observables studied there, the problematic multi-edge correlations are absent.

First-pass degeneracy is not the primary cause: exhaustive enumeration at $d=3$ and $d=5$ shows that most size-$\frac{d-1}{2}$ logical failures persist even when the first-pass correction exactly matches the sampled errors.

\textit{Regularized correlated decoding.}---   
To recover the correlated decoders' effective distance, we introduce a regularization parameter $\alpha$, which we refer to as the reweighting strength, into the reweighting rule of Eq.~\eqref{eq:CM}:
\begin{equation}
    p'_\nu(\alpha) = \alpha P(e_\nu\mid e_\mu).
\end{equation}

The regularization adds $\sim\log(1/\alpha)$ to every discounted weight, so a correction that uses a first-pass edge multiple times pays a higher penalty, as shown in the Appendix, Eq.~\eqref{eq:dw-dec}.
At high $p$, $\PL$ is dominated by fault configurations larger than $\frac{d-1}{2}$, which the standard correlated update corrects effectively. 
This favors strong reweighting near threshold. 
The low-$p$ regime requires more moderate discounts, so we interpolate between both limits via an empirical model for the optimal regularization strength $\astar$:
\begin{equation}
    \astar(p)=\frac{\alpha_0}{1+(p_0/p)^k},
    \label{eq:alphastar}
\end{equation}
where $\alpha_0$, $p_0$, and $k$ are fitted constants. 
We find a good agreement between the model and our data for both MWPM and LCD as shown in Section~\ref{sm:regularization-fits} of the Appendix.
Alternative models, potentially derived from the specific failure modes present in a given circuit, may also be suitable.

\begin{figure}[!ht]
    \centering
    \includegraphics[width=\columnwidth]{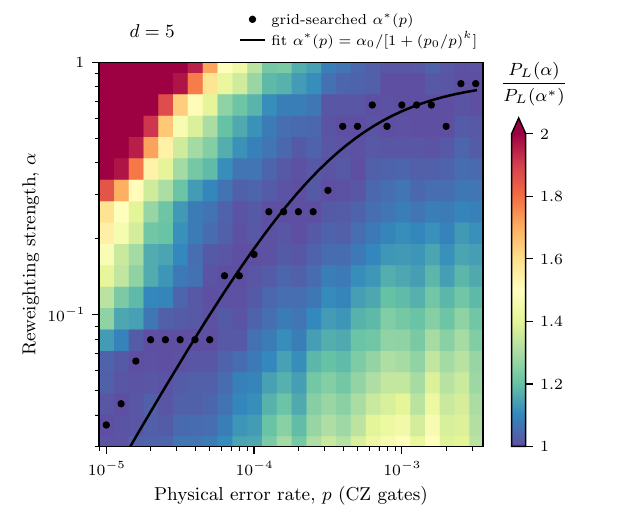}
    \caption{\textbf{Error-rate-dependent regularization tracks the
    optimal reweighting ridge.}
    Logical-error landscape of RCM at $d=5$ under SI1000 noise with $r=d$.
    The color scale shows $\PL(\alpha)/\min_{\tilde\alpha}\PL(\tilde\alpha)$ over the tested strengths at each $p$. Black circles mark those empirical minimizers; the solid curve is the interpolation $\astar(p)$ in Eq.~\eqref{eq:alphastar}.
    Standard correlated matching corresponds to $\alpha=1$.
    Near threshold the optimum remains close to the standard update,
    whereas deep below threshold it shifts toward weaker cross-subgraph
    reweighting.}
    \label{fig:alpha-landscape}
\end{figure}

To test the effectiveness of the regularization and the validity of the model for the optimal reweighting strength $\astar$ of Eq.~\eqref{eq:alphastar}, we perform a grid search over values of $\alpha$ for a range of physical error rates, shown in Figure~\ref{fig:alpha-landscape}.
For each $(p,\alpha)$ pair, we calculate the logical error probability ratio of each regularized decoder to its uncorrelated variant and identify the value $\astar$ that minimizes it. 
Given the optimal $\astar(p)$, we refer to the resulting decoders as Regularized Correlated Matching (RCM) and Regularized Correlated LCD (RC-LCD).
The solid line in Figure~\ref{fig:alpha-landscape} depicts the optimal fitted $\astar$ for the matching decoder on a $d=5$ rotated surface code, as given by Eq.~\eqref{eq:alphastar}, placed on top of the full grid search of candidate $\alpha$-values.
We find a smooth ridge of favorable regularization strengths, with the empirical optimum $\astar$ near unity at higher $p$ and decreasing as $p$ is reduced.
The full sweep details, including $\astar$ fits for RCM and RC-LCD across $d=\{3, 5, 7, 9\}$, are given in Section~\ref{sm:regularization-fits} of the Appendix.
Both RCM and RC-LCD show similarly broad valleys around $\astar$, suggesting that the optimal regularization strength depends weakly on the decoder when the reweighting protocol is fixed. 

Given an optimal $\astar(p)$, we compare the logical performance of RCM and RC-LCD against the standard two-pass decoder variants in Fig.~\ref{fig:degradation}. 
The accuracy reversal is fully removed for both regularized decoders, which outperform both the uncorrelated and the standard correlated decoders down to the lowest $p$ sampled.
At $p=10^{-4}$, the RCM-to-matching ratios are $\sim0.55$ at $d=5$ and $\sim0.35$ at $d=7$, compared with $\sim0.73$ and $\sim0.45$ for standard CM. 
For LCD at $d=5$ and $p=5\times10^{-5}$, regularization reduces the correlated-to-uncorrelated ratio from $\sim0.95$ to $\sim0.55$.
At $p\sim10^{-5}$, $\astar(p)\simeq0.03$ remains nonzero, while the regularized-to-uncorrelated error-rate ratios approach constants below unity, showing that regularization preserves a stable benefit from correlations in the deep sub-threshold regime.

In addition to the markedly improved logical error rates, we show that the minimum-size failure contribution is significantly reduced with regularization.
As shown in Fig.~\ref{fig:low_size_p_L_contrib}, RCM eliminates the single-fault contribution at $d=3$. At $d=5$ and $d=7$, it strongly suppresses the size-$\frac{d-1}{2}$ contribution relative to CM, keeping it more than an order of magnitude below the size-$\frac{d+1}{2}$ contribution (Fig.~\ref{fig:low_size_p_L_contrib}). 
In addition, RCM maintains CM's accuracy for decoding size-$\frac{d+1}{2}$ errors, outperforming MWPM.
These results illustrate how regularization preserves useful correlation information while strongly suppressing the minimum-size failures that were responsible for the low-$p$ breakdown.

\textit{Discussion and outlook.}---
Standard correlation reweighting can introduce unexpected failures in typically correctable fault configurations deep below threshold. 
These failures arise because the reweighting rule discounts several edges conditioned on the same first-pass edge without accounting for the additional independent faults required for them to occur jointly. This can favor a correction requiring more faults than the physical error and allow configurations containing $\frac{d-1}{2}$ faults to trigger logical failures. 
Regularization dampens the overly aggressive reweighting protocol and substantially suppresses the observed failures while retaining the accuracy benefit of correlations across the error rates studied.
The regularized updates can be precomputed at initialization, leaving the underlying decoding routines of clustering and matching unaffected and ensuring that our approach remains as computationally efficient as the standard correlated decoding protocols.

The single-parameter regularization studied here provides a simple remedy for the observed breakdown, while leaving open the broader problem of how to incorporate correlations without sacrificing low-error performance.
Our results motivate a new design principle not only for correlated decoders but, more generally, for any decoder that uses conditional error-model updates: the combined cost of discounted edges should retain the probability penalties for any additional independent faults required to explain them jointly. 
This principle offers a starting point for improving the update protocols in a range of high-accuracy decoding techniques, such as belief-matching~\cite{higgott2023improved,caune2023belief}, soft-information-aware decoders~\cite{pattison2021improved, majaniemi2026reducing}, and leakage-aware decoders~\cite{ziad2025local, Bonilla_Ataides_2026}.

If graph weights and updates are learned using machine learning, our analysis may help interpret departures from standard correlation reweighting even when regularization is not imposed explicitly. 
The regularized updates could also provide a physically motivated prior for learning such refinements.
Investigating analogous constraints in belief-propagation decoders for qLDPC codes~\cite{muller2025improved} would test the scope of this approach and may guide the development of accurate, low-latency correlated decoding beyond surface codes. 
These questions are particularly relevant to hierarchical architectures~\cite{gidney2023yokedsurfacecodes, pattison2023hierarchical, Chamberland_2023}, where error suppression by an inner code exposes the outer decoder to very low effective error rates, making performance deep below threshold central to the reliability of the full architecture.

\section*{Acknowledgments}
We thank Oscar Higgott for useful discussions on hyperedge decomposition, Barbara Terhal, Mark Turner, Mark Walters, Matthew Stafford, and other members of Riverlane for insightful discussions, and the Applied Quantum Algorithms group at Leiden University for valuable feedback. 
Claude Code with the Opus 5 model was used to assist with derivations, writing code for simulations, data analysis, and plotting -- all output was reviewed and verified by the authors.

\section*{Data and Code Availability}
The data, matching-decoder code, and analysis scripts supporting this work are available in the \href{https://github.com/cesar-hernando/regularized-correlated-matching}{\texttt{regularized-correlated-matching}} repository. 
The LCD implementation is proprietary to Riverlane.

\bibliographystyle{apsrev4-2}
\bibliography{references}
\clearpage
\appendix
\setcounter{secnumdepth}{2}

\section{Logical error rate scaling}

The expected behavior of the logical error rate $P_\textrm{L}$ of a distance-$d$ error-correction code at low physical error rates $p$ is:
\begin{equation}
    P_\textrm{L}\propto p^{(d+1)/2}
    \label{eq:pL_vs_d}
\end{equation}
where errors of size $(d+1)/2$ are those most likely to dominate the logical error budget~\cite{fowler2012surface}.
Figure~\ref{fig:lep_vs_p} compares the logical error probabilities of the uncorrelated, correlated, and regularized matching and clustering decoders under SI1000 noise. 
The standard one-pass (uncorrelated) decoder meets the nominal logical error scaling of Eq.~\eqref{eq:pL_vs_d}, while the correlated decoder with $\alpha=1$ deviates from the predicted error rate scaling in the deep-subthreshold error regime.
Table~\ref{tab:exponents} gives the slope of $\log \PL$ against $\log p$ over a low-$p$ window.
The regularized correlated decoder achieves the nominal error suppression exponent $\gamma=(d+1)/2$ or higher at every distance sampled, while the standard decoder shows worse-than-expected scaling at low $p$.
This difference in error suppression makes regularization essential for achieving low logical error rates and directly affects the required physical qubit footprint of a fault-tolerant device.

\section{Effective-distance diagnostics}
\label{app:distance}

For odd code distance $d$, let $t=(d-1)/2$.
A decoder that preserves the nominal distance corrects every configuration of at most $t$ faults, giving
\begin{equation}
    \PL
    =
    A p^{t+1}
    +
    \Ord(p^{t+2}).
    \label{eq:distance-preserving-scaling}
\end{equation}
We test this condition directly by decomposing the logical error probability into contributions $\PL^{(m)}$ from configurations containing exactly $m$ error mechanisms, as found in the circuit detector error model. 
These contributions are probability-weighted since different SI1000 error mechanisms have different leading coefficients.
A persistent contribution from the $m=t$ sector therefore identifies logical failures caused by nominally correctable fault configurations, independently of fits to the aggregate logical error probability.
An order-$p^t$ term persists under a changed parametrization if at least one such configuration retains a probability proportional to $p^t$ and remains a decoder failure as $p\to0$.

Fig.~\ref{fig:low_size_p_L_contrib} shows how the probability-weighted contributions of configurations with exactly $t$ and $t+1$ faults differ between standard matching, correlated matching, and regularized correlated matching.
Let $\pi_i$ denote the probability of independent DEM mechanism $i$.
Each contribution is the total probability of failing configurations containing exactly $m$ mechanisms
\begin{equation}
    \PL^{(m)}
    =
    \sum_{S}\prod_{i\in S}\pi_i
    \prod_{j\notin S}(1-\pi_j),
\end{equation}
where the sum runs over all sets $S$ of $m$ distinct DEM mechanisms on which the decoder fails. 
At $d=3$ and $d=5$, we evaluate this sum exactly by decoding every set.
At $d=7$, the sets are too numerous to enumerate (of order $10^{10}$ for $m=3$ and $10^{14}$ for $m=4$), so we estimate the sums by uniformly sampling $N$ sets of $m$ distinct mechanisms, decoding each, and estimating
\begin{equation}
    \PL^{(m)}
    \simeq
    \binom{M}{m}N^{-1}
    \sum_{S}\prod_{i\in S}\pi_i
    \prod_{j\notin S}(1-\pi_j).
\end{equation}
Here $M$ is the total number of DEM mechanisms, and the sum runs over the sampled sets $S$ on which the decoder fails. 
We use $N=6.4\times10^{8}$ samples for $m=3$ and $N=3.2\times10^{8}$ for $m=4$ at each error rate, and the error bars are the standard errors of these sample means.

\onecolumngrid
\vfill
\noindent
\begin{minipage}{\textwidth}
\makeatletter
\def\@captype{table}
\makeatother
\small
\centering
\caption{Fitted error-suppression exponents $\gamma$ for a scaling law of the form $\PL\propto p^\gamma$ for MWPM and LCD given a fixed window in the error rate $p$. 
The error probabilities are estimated from $d$-round rotated planar memory experiments under SI1000 noise by taking a $\log$-$\log$ fit from the six lowest-$p$ resolved data points. 
Results that deviate significantly from the expected scaling of the one-pass decoder are highlighted in bold. 
The $d=9$ samples have a higher sampling window in $p$, leading to a smaller deviation compared to the baseline.}
\label{tab:exponents}
\renewcommand{\arraystretch}{1.3}
\begin{tabular*}{\textwidth}{@{\extracolsep{\fill}}llcccccc@{}}
\toprule
Decoder & $d$ & $p_{\min}$ & $p_{\max}$ & $\gamma$ (one-pass) & $\gamma$ (two-pass) & $\gamma$ $(\astar)$ & Ideal $(d+1)/2$\\
\midrule
MWPM  & 3 & $1.0\times10^{-5}$ & $3.2\times10^{-5}$ & $1.98 \pm 0.05$ & $\mathbf{1.28 \pm 0.03}$ & $1.97 \pm 0.05$ & 2 \\
MWPM  & 5 & $1.0\times10^{-5}$ & $3.2\times10^{-5}$ & $3.02 \pm 0.05$ & $\mathbf{2.25 \pm 0.04}$ & $2.95 \pm 0.06$ & 3 \\
MWPM  & 7 & $6.3\times10^{-5}$ & $2.0\times10^{-4}$ & $4.04 \pm 0.05$ & $\mathbf{3.74 \pm 0.07}$ & $3.99 \pm 0.09$ & 4 \\
MWPM  & 9 & $2.0\times10^{-4}$ & $6.3\times10^{-4}$ & $5.10 \pm 0.07$ & $5.21 \pm 0.15$ & $5.36 \pm 0.16$ & 5 \\
\midrule
LCD & 3 & $1.0\times10^{-5}$ & $3.2\times10^{-5}$ & $1.97 \pm 0.03$ & $\mathbf{1.30 \pm 0.02}$ & $1.95 \pm 0.03$ & 2 \\
LCD & 5 & $2.0\times10^{-5}$ & $6.3\times10^{-5}$ & $3.01 \pm 0.04$ & $\mathbf{2.48 \pm 0.04}$ & $3.01 \pm 0.05$ & 3 \\
LCD & 7 & $1.0\times10^{-4}$ & $3.2\times10^{-4}$ & $4.04 \pm 0.04$ & $\mathbf{3.82 \pm 0.06}$ & $4.02 \pm 0.07$ & 4 \\
LCD & 9 & $2.5\times10^{-4}$ & $7.9\times10^{-4}$ & $5.16 \pm 0.04$ & $5.22 \pm 0.08$ & $5.37 \pm 0.08$ & 5 \\
\bottomrule
\end{tabular*}
\end{minipage}
\par
\clearpage
\onecolumngrid
\noindent
\begin{minipage}{\textwidth}
\makeatletter
\def\@captype{figure}
\makeatother
    \centering
    \includegraphics[width=\linewidth]{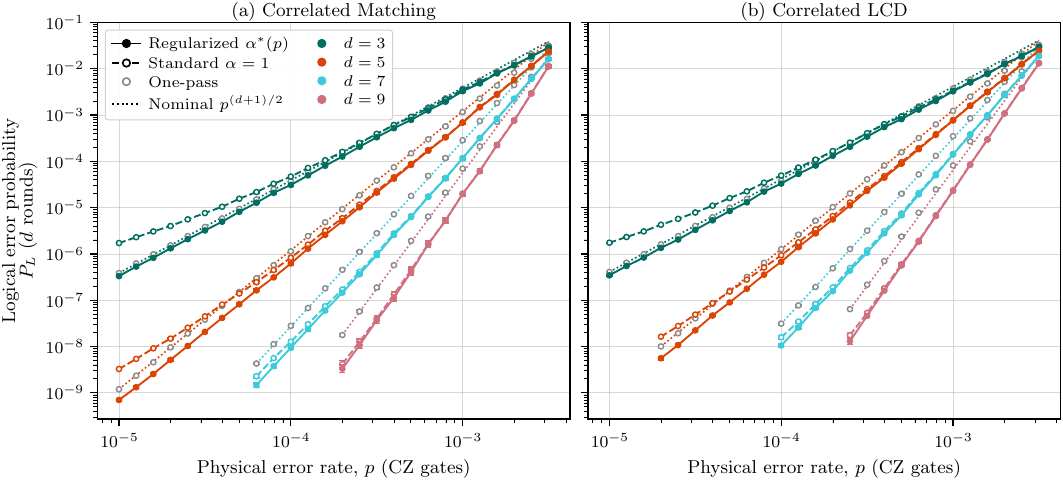}
    \caption{\textbf{Logical error probability $P_\textrm{L}$ for $d$-round memory as a function of physical error rate $p$ under SI1000 noise, decoded with MWPM and LCD}. 
    The logical performance is evaluated for the uncorrelated (one-pass) decoder, the standard two-pass decoder ($\alpha=1$) and the optimal regularized correlated decoder with $\alpha=\astar$. 
    Each point is sampled using \textsc{Stim}. For LCD, sampling stops after $4000$ failures or $1.2\times10^{11}$ shots. 
    For MWPM, each point uses a fixed number of shots (up to $1.9\times10^{12}$), chosen to match the LCD shot count or to target approximately $2000$ MWPM failures.
    Error bars are plotted, but are negligible for the majority of the data points. 
    The non-regularized decoder diverges from the expected scaling law and has increasingly higher error rates relative to the uncorrelated decoder as $p\rightarrow0$, while the regularized decoder consistently outperforms the uncorrelated decoder and recovers the $p^{(d+1)/2}$-scaling.
    }
    \label{fig:lep_vs_p}
\end{minipage}
\par
\medskip
\twocolumngrid

For MWPM, the $t$-fault contribution is either absent or confined to large $p$, where it falls off steeply as $p$ decreases.
Standard CM instead produces a $t$-fault contribution that decays only as $p^{t}$, one power of $p$ below the distance-preserving scaling.
At $d=3$ this contribution already exceeds the $(t+1)$-fault contribution at the lowest error rates shown, and at $d=5$ and $d=7$ the gap between the two narrows as $p$ decreases.
RCM removes the $t$-fault contribution entirely at $d=3$ and, at $d=5$ and $d=7$, keeps it below that of CM with a steeper decay toward low $p$, more than an order of magnitude below its own $(t+1)$-fault contribution.
RCM also handles $(t+1)$-fault configurations about as well as CM, so although its reweighting weakens at lower $p$, it still benefits from the correlation information.

\section{Decision condition for a correlated decoder}
\label{app:decision}

Equations~\eqref{eq:joint-true}--\eqref{eq:joint-assumed} translate directly into the decoder's choice between corrections.
Given a true partner edge $e_k$ for the first-pass edge $e_\mu$ (\textit{true partner} meaning that $e_k$ and $e_\mu$ are part of the same hyperedge mechanism that occurred), we denote by $C_\checkmark$ the correct second-pass correction string containing $e_k$, and by $C_\times$ a correction with the opposite logical effect, containing both partner edges $e_\nu, e_\sigma$ instead.
Discounted edges have weights $w'$ as per Eq.~\eqref{eq:CM} (with $\alpha q$ in place of $q$ when regularized), while every other edge has weight $\log(1/p)+\Ord(1)$.
The decoder compares the two correction weights of the two paths and observes a difference $\Delta w_{\rm dec}$ given by:
\begin{equation}
\begin{split}
    &\Delta w_{\rm dec} \equiv w_{\rm dec}(C_\times)-w_{\rm dec}(C_\checkmark)\\
    &\simeq \Bigl(n_\times \log\tfrac{1}{p}+w'_\nu+w'_\sigma\Bigr)
          - \Bigl(n_\checkmark \log\tfrac{1}{p}+w'_k\Bigr)\\
    &= \Delta n\log\tfrac1p + \bigl(w'_\nu + w'_\sigma - w'_k\bigr),
\end{split}
\label{eq:dw-dec}
\end{equation}
where $n_\times$ and $n_\checkmark$ count the baseline (non-reweighted) edges in the competing and correct paths, respectively, and $\Delta n=n_\times-n_\checkmark$. 
The true cost difference $\Delta w_{\rm true}$ instead follows from
Eq.~\eqref{eq:joint-true}. As only one of $e_\nu, e_\sigma$ can be explained by a
mechanism containing $e_\mu$, the other requires an independent fault of probability $r=\Ord(p)$:
\begin{equation}
\begin{split}
    &\Delta w_{\rm true} \equiv w_{\rm true}(C_\times)-w_{\rm true}(C_\checkmark)\\
    &\simeq \Bigl(n_\times\log\tfrac{1}{p}-\log\bigl[q_\nu r_\sigma+q_\sigma r_\nu\bigr]\Bigr) - \Bigl(n_\checkmark\log\tfrac{1}{p}+\log\tfrac{1}{q_k}\Bigr)\\
    &= (\Delta n) \log \tfrac{1}{p}+\log \frac{q_k}{q_\nu r_\sigma+q_\sigma r_\nu}\\
    &= (\Delta n+1)\log\tfrac{1}{p} + \log\frac{q_k}{q_\nu\rho_\sigma+q_\sigma\rho_\nu} + \Ord(p)\\
    &= (\Delta n+1)\log\tfrac{1}{p} + \Ord(1).
\end{split}
\label{eq:dw-true}
\end{equation}

\clearpage
\onecolumngrid
\noindent
\begin{minipage}{\textwidth}
\makeatletter
\def\@captype{figure}
\makeatother
    \centering
    \includegraphics[width=\textwidth]{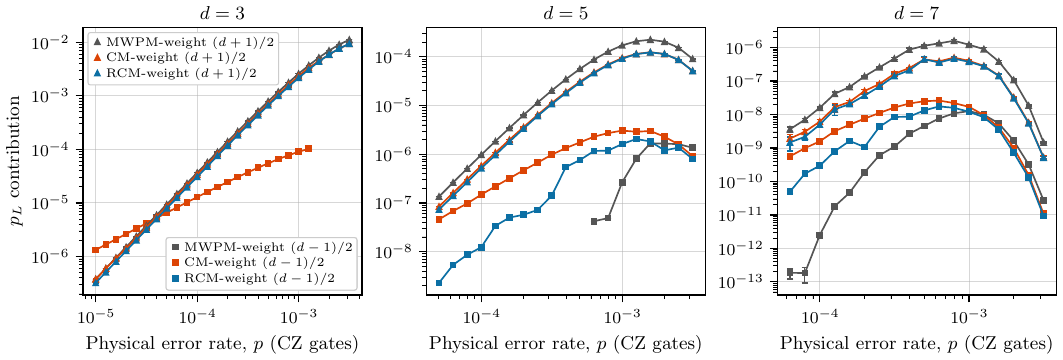}
    \caption{\textbf{Probability-weighted contributions to $\PL$ from fixed-size errors.} Results are shown for rotated planar code memory simulations under SI1000 noise with $r=d$ syndrome-extraction rounds and code distances $d=\{3,5,7\}$. We evaluate contributions from configurations containing exactly $t=(d-1)/2$ faults (squares) and $t+1=(d+1)/2$ faults (triangles), resolved for matching, correlated matching (CM), and regularized correlated matching (RCM).
    Persistence of the $t$-fault contribution toward low $p$ diagnoses a loss of the nominal distance scaling.
    Standard CM generates such a contribution, while RCM suppresses it.
    }
    \label{fig:low_size_p_L_contrib}
\end{minipage}
\par
\medskip
\twocolumngrid

Here $r_\sigma=\rho_\sigma p+\Ord(p^2)$, with $\rho_\sigma$ as a $p$-independent constant that describes the rate at which $e_\sigma$ is produced by mechanisms not containing $e_\mu$. 
The standard rule thus misses a contribution of $\log(1/p)$ for each reused source edge, making such faults appear free.
When $\Delta n=0$, as is the case in the $d=3$ example of the main text where both corrections consist only of discounted edges, $C_\times$ is chosen whenever $q_\nu q_\sigma>q_k$, independent of $p$. 
This makes the decoder choose the wrong path whenever the true partner edge $e_k$ is rarer than the pair of its competitors $e_\nu, e_\sigma$.
Regularization adds $\log(1/\alpha)$ to $\Delta w_{\rm dec}$, preventing the $\Delta n=0$ failures when $\alpha<q_k/(q_\nu q_\sigma)$.

\section{Simulation and statistics}
\label{sm:simulation_details}

\subsection{Circuit construction}
\label{sm:circuit_construction}

\begin{figure}[hb]
    \centering
    \includegraphics[width=1.0\linewidth]{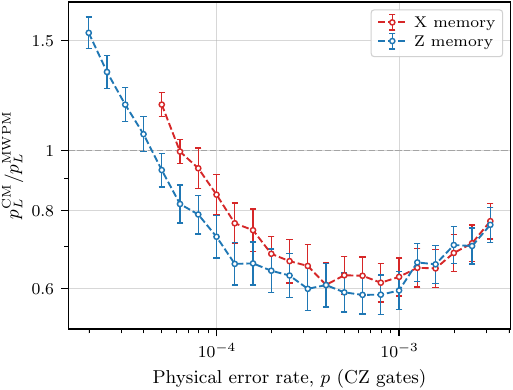}
    \caption{\textbf{$X$- and $Z$-basis logical memory error probability ratios for correlated vs uncorrelated MWPM under varying error rates $p$.} The data is obtained from a simulated $d$-round quantum memory experiment under SI1000 noise at code distance $5$. 
    Error bars combine individual binomial likelihood intervals under an independence approximation; see Appendix~\ref{sm:uncertainty}. Both the $X$- and $Z$-basis memory experiments show the characteristic correlated decoder degradation at low $p$. The effect is stronger for $X$-memory, likely due to asymmetry in the $X$- vs $Z$-stabilizer syndrome extraction circuits.}
    \label{fig:x_vs_z_degradation}
\end{figure}

We use rotated planar-code memory circuits of distances $d\in\{3,5,7,9\}$ with $r=d$ syndrome-extraction rounds followed by data-qubit measurement.
Every reset, measurement, idle time, and one- or two-qubit gate is assigned its SI1000 error channel~\cite{gidney2021fault}.
The parameter $p$ is the two-qubit depolarizing error probability; all other channel probabilities follow the SI1000 prescription.
We use $26$ logarithmically spaced values of $p$ from $10^{-5}$ to $3.16\times10^{-3}$.
The main results use $Z$-memory circuits; Figure~\ref{fig:x_vs_z_degradation} shows that the degradation also occurs in $X$-memory circuits. 
The degradation for $X$-memory appears stronger than for $Z$-memory, making our results a \textit{conservative} estimate of the true impact of correlated decoding failure modes.
To sample circuits, we use the fast Clifford simulator \textsc{Stim}~\cite{gidney2021stim}, first converting the circuits into detector error models (DEMs) and then sampling independent error mechanisms from each DEM using the \texttt{compile\_detector\_sampler} factory.
To make the raw DEM decodable by MWPM, we decompose hyperedges into graphlike components using $X$--$Z$ annotations from the \textsc{Tesseract} repository~\cite{beni2025tesseract}. The LCD results are based on the native \textsc{Stim} decomposition---the difference between the \textsc{Tesseract} method and native \textsc{Stim} was statistically insignificant.

\subsection{Results under uniform depolarizing noise}
\label{sm:uniform_dep_noise}

\begin{figure}[h]
    \centering
    \includegraphics[width=1\linewidth, left]{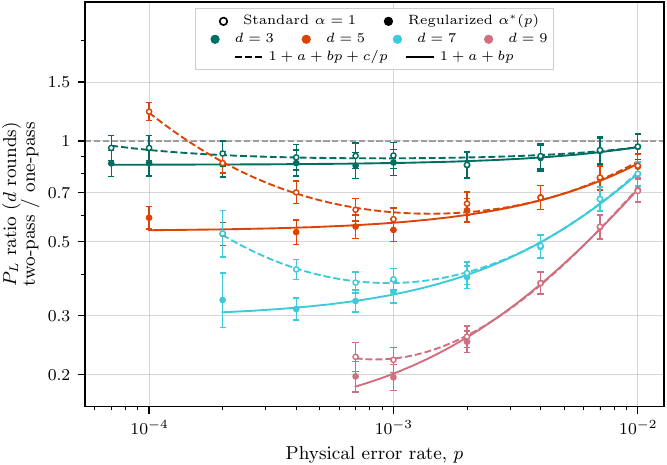}
    \caption{\textbf{Ratio of logical error probabilities for standard and regularized correlated matching with respect to uncorrelated MWPM under uniform depolarizing noise.} Open markers denote the standard update, $\alpha=1$, and filled markers the regularized update with $\alpha^*(p)$. Dashed curves show fits of the form $1+a+bp+c/p$, and solid curves fits of the form $1+a+bp$. The horizontal line marks equal performance. As $p\rightarrow0$, the standard correlated decoder degrades relative to the uncorrelated decoder, while regularization restores a correlated decoding advantage across the error rates studied.}
    \label{fig:uniform_rcm}
\end{figure}

To confirm that the accuracy reversal of the correlated decoders at low $p$ is not an artifact of the SI1000 noise model used, we ran a sweep of the correlated matching decoder under uniform depolarizing noise, where every circuit-level operation is subject to depolarizing errors with probability $p$.
The result is shown in Fig.~\ref{fig:uniform_rcm}.
Under the uniform noise model, we see the same characteristic reversal of the correlated decoding error rates at low $p$, with the distance-$5$ code showing parity between the correlated and uncorrelated matching decoders at $p\sim1.4\times10^{-4}$. 
Regularization restores the accuracy advantage of correlated decoding and results in a constant logical error rate ratio at low $p$, similar to what was observed for the SI1000 model.

\subsection{Uncertainty analysis}
\label{sm:uncertainty}

For each decoder, we estimate the logical error probability
as $\widehat{\PL}=E/N$, where $E$ is the number of logical
failures among $N$ shots.
We obtain individual binomial likelihood intervals using
\textsc{Sinter} with maximum likelihood factor
$10$~\cite{Gidney2024StackOverflow}, equivalent to a $96.8\%$
two-sided Gaussian interval.
For ratios of logical error probabilities, we combine the
one-sided logarithmic deviations of the numerator and
denominator in quadrature, treating their estimates as
independent.
This approximation neglects the positive covariance arising from
decoding the same syndrome samples, so the resulting intervals
are conservative.

The number of shots taken in our simulations to estimate the ratio of logical error rates between the correlated and uncorrelated decoders varies based on code distance---we set a target minimum of $n=4000$ errors for the local clustering decoder and $n=2000$ for the minimum-weight perfect matching decoder.
In cases where sampling this many errors is unfeasible, such as when $p\rightarrow10^{-4}$ at distances $d=7$ and above, we take up to 
$1.9\times10^{12}$ shots for the most expensive data points for correlated MWPM and $1.2\times10^{11}$ for correlated LCD.
The full dataset of decoder failures, the upper and lower bounds on the logical error rate ratios, and the number of shots taken can be viewed in the \href{https://github.com/cesar-hernando/regularized-correlated-matching}{\texttt{regularized-correlated-matching}} repository. A fork of \textsc{PyMatching} that enables regularized correlated decoding can be found at \href{https://github.com/cesar-hernando/PyMatching/tree/regularized_cm}{\texttt{PyMatching-v2.3.1+rcm.1}}.

\section{Weight scales and ratio fits}
\label{sm:weight_scales_and_regularization}

\subsection{Posterior update rule for correlations}
\label{sm:weight-scales}

\begin{figure*}
    \centering
    \includegraphics[width=1.0\textwidth]{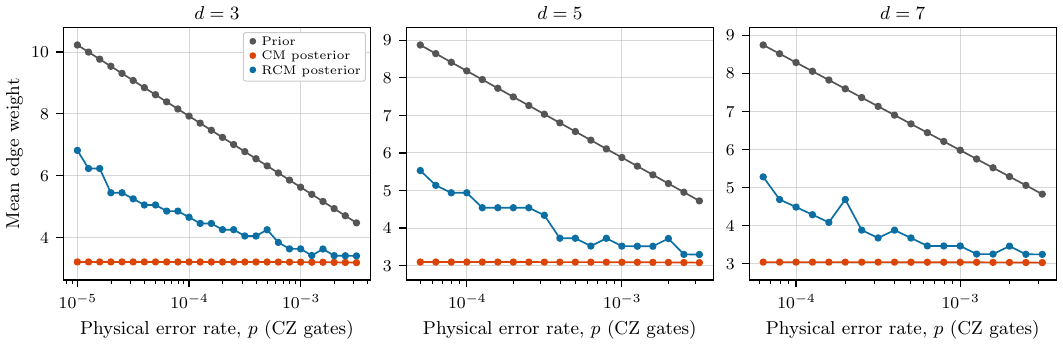}
    \caption{\textbf{Mean edge weights of priors and CM posteriors with and without regularization.} Results are shown for rotated planar code memory simulations under SI1000 noise with $r=d$ syndrome-extraction rounds and code distances $d=\{3,5,7\}$. 
    Mean prior edge weights are shown in gray, mean edge weights after standard correlated matching (CM) are shown in orange, and the regularized correlated matching (RCM) weights are shown in blue. 
    The CM weights are $\Ord(1)$ regardless of the error rate $p$, while regularized correlated matching introduces a $p$-dependence similar to, but weaker than, that of the original graph priors.
    }
    \label{fig:mean_weights}
\end{figure*}

An edge is activated when an odd number of DEM mechanisms containing it occur.
Let $\pi_h$ denote the probability of independent DEM mechanism $h$,
and let $H_\nu$ contain the mechanisms involving edge $e_\nu$.
The marginal activation probability of $e_\nu$ is
\begin{equation}
    p_\nu
    =
    \frac{1-\prod_{h\in H_\nu}(1-2\pi_h)}{2}
    =
    c_\nu p+\Ord(p^2).
\end{equation}
Consequently,
\begin{equation}
    w_\nu
    =
    \log\frac{1-p_\nu}{p_\nu}
    =
    \log\frac1p-\log c_\nu+\Ord(p).
\end{equation}
If $e_\mu$ and $e_\nu$ share a first-order fault,
\begin{equation}
    P(e_\mu\wedge e_\nu)
    =
    c_{\mu\nu}p+\Ord(p^2),
    \qquad
    P(e_\mu)=c_\mu p+\Ord(p^2).
\end{equation}
With
\begin{equation}
    q_{\nu\mid\mu}
    =
    \min\!\left\{\frac{c_{\mu\nu}}{c_\mu},\frac12\right\},
\end{equation}
the candidate update conditioned on a selected edge $e_\mu$ satisfies
\begin{equation}
    p'_\nu=q_{\nu\mid\mu}+\Ord(p),
    \qquad
    w'_\nu=\Ord(1).
\end{equation}
CM takes the largest capped posterior probability over all selected
correlated edges in the opposite subgraph, replacing the prior only
when this probability exceeds $p_\nu$.
Equivalently, it selects the minimum of the prior weight and all
candidate updated weights.
Thus $w'_\nu/w_\nu\to0$ as $p\to0$.
This separation of scales can cause the decoder to underestimate the cost of corrections that combine several edges discounted by the same source edge, because their joint occurrence may require additional independent faults. 
Figure~\ref{fig:mean_weights} shows the mean weights for edges in rotated planar surface code decoding graphs under SI1000 noise, highlighting differences in the $p$-dependence of prior, reweighted, and regularized edge weights.

\subsection{Logical error ratio fits}
\label{sm:ratio-fits}
In this subsection, we derive the ansatz used to fit the logical error probability ratios of the standard and regularized correlated decoders with respect to the uncorrelated ones in Fig.~\ref{fig:degradation}.
First, we note that the uncorrelated decoders preserve the code distance, meaning that the lowest number of independent circuit faults that generate a logical error is $\frac{d+1}{2}$. 
Thus, we expand the logical error probability of uncorrelated or independent decoders (e.g., MWPM and LCD) as
 \begin{equation}
    \PL^{\rm ind}=Ap^\frac{d+1}{2}+\Ord(p^\frac{d+3}{2}), 
    \label{eq:ind_p}
 \end{equation}
 where $A$ is a distance-dependent factor. 
 In contrast, the standard correlated decoders suffer a loss of one unit of effective distance, as they can incorrectly decode size-$\frac{d-1}{2}$ fault configurations that a distance-$d$ code should be resilient to. 
 Thus, their logical error probability can be expanded as:
 \begin{equation}
     \PL^{\rm corr}=Bp^\frac{d-1}{2}+Cp^\frac{d+1}{2}+Dp^\frac{d+3}{2} + \Ord(p^{\frac{d+5}{2}}), 
     \label{eq:cm_p}
 \end{equation}
 where $B$, $C$, and $D$ are also distance-dependent parameters. Dividing Eq.~\eqref{eq:cm_p} by Eq.~\eqref{eq:ind_p}, we obtain the ratio of the logical error probabilities of the correlated and independent decoders:
 \begin{equation}
     \frac{\PL^{\rm corr}}{\PL^{\rm ind}}=\frac{B}{Ap}+\frac{C}{A}+\frac{D}{A}p+\Ord(p^2).
 \end{equation}
 By renaming the parameters $C/A=1+a$, $B/A=c$, $D/A=b$, and ignoring quadratic and higher terms, we obtain the model used to fit the standard correlated decoder data points of Fig.~\ref{fig:degradation}:
 \begin{equation}
     \frac{\PL^{\rm corr}}{\PL^{\rm ind}}=1+a+bp+\frac{c}{p}.
     \label{eq:cm_ratio_model}
 \end{equation}

Regularized correlated decoders show only a negligible loss of effective distance (Figs.~\ref{fig:low_size_p_L_contrib} and \ref{fig:lep_vs_p}) while retaining the accuracy of standard correlated decoding on larger fault configurations.
 Thus, we can suppress the $c/p$ term related to the effective distance loss in Eq.~\eqref{eq:cm_ratio_model}, and obtain the ansatz used to fit the logical error probability ratios for the regularized correlated decoders in Fig.~\ref{fig:degradation}:
 \begin{equation}
     \frac{\PL^{\alpha^*{\rm-corr}}}{\PL^{\rm ind}}=1+a+bp.
 \end{equation}

\begin{table}[h]\centering
\caption{Fit parameters for the ratio of standard two-pass ($\alpha=1$) to one-pass logical error probabilities:}
\par\smallskip
$\PL^{\mathrm{CM}}/\PL^{\mathrm{MWPM}}~(\PL^{\mathrm{C-LCD}}/\PL^{\mathrm{LCD}})= 1+a+bp+c/p$
\par\smallskip
\vspace{2pt}
\begin{tabular}{llcccr}
\toprule
decoder & $d$ & $a$ & $b$ & $c$ & $R^2$ \\
\midrule
CM & 3 & $-0.09 \pm 0.01$ & $-10 \pm 8$ & $(3.48 \pm 0.06)\times10^{-5}$ & 0.999 \\
CM & 5 & $-0.489 \pm 0.009$ & $79 \pm 7$ & $(2.08 \pm 0.06)\times10^{-5}$ & 0.996 \\
CM & 7 & $-0.73 \pm 0.01$ & $114 \pm 7$ & $(1.8 \pm 0.2)\times10^{-5}$ & 0.983 \\
CM & 9 & $-0.85 \pm 0.02$ & $112 \pm 11$ & $(1.2 \pm 0.7)\times10^{-5}$ & 0.972 \\
\addlinespace
C-LCD & 3 & $-0.090 \pm 0.007$ & $-10 \pm 5$ & $(3.44 \pm 0.03)\times10^{-5}$ & 0.999 \\
C-LCD & 5 & $-0.479 \pm 0.006$ & $76 \pm 4$ & $(2.13 \pm 0.05)\times10^{-5}$ & 0.996 \\
C-LCD & 7 & $-0.709 \pm 0.008$ & $109 \pm 5$ & $(1.8 \pm 0.2)\times10^{-5}$ & 0.982 \\
C-LCD & 9 & $-0.84 \pm 0.01$ & $108 \pm 6$ & $(1.9 \pm 0.5)\times10^{-5}$ & 0.992 \\
\addlinespace
\bottomrule
\end{tabular}
\end{table}

\begin{table}[h]\centering
\caption{Fit parameters for the ratio of regularized two-pass ($\astar$) to one-pass logical error probabilities:}
\par\smallskip
$\PL^{\mathrm{RCM}}/\PL^{\mathrm{MWPM}}~(\PL^{\mathrm{RC-LCD}}/\PL^{\mathrm{LCD}}) = 1+a+bp$
\par\smallskip
\vspace{2pt}
\begin{tabular*}{\columnwidth}{@{\extracolsep{\fill}}llccr@{}}
\toprule
decoder & $d$ & $a$ & $b$ & $R^2$ \\
\midrule
RCM & 3 & $-0.151 \pm 0.006$ & $20 \pm 6$ & 0.659 \\
RCM & 5 & $-0.464 \pm 0.005$ & $68 \pm 5$ & 0.969 \\
RCM & 7 & $-0.693 \pm 0.006$ & $98 \pm 5$ & 0.976 \\
RCM & 9 & $-0.838 \pm 0.008$ & $107 \pm 6$ & 0.978 \\
\addlinespace
RC-LCD & 3 & $-0.152 \pm 0.004$ & $21 \pm 4$ & 0.802 \\
RC-LCD & 5 & $-0.462 \pm 0.003$ & $69 \pm 3$ & 0.964 \\
RC-LCD & 7 & $-0.691 \pm 0.004$ & $101 \pm 3$ & 0.986 \\
RC-LCD & 9 & $-0.816 \pm 0.004$ & $98 \pm 3$ & 0.996 \\
\addlinespace
\bottomrule
\end{tabular*}
\end{table}

\section{Representative failure modes}
\label{sm:failure_examples}

\begin{figure*}[p]
    \centering
    \includegraphics[height=0.9\textheight]{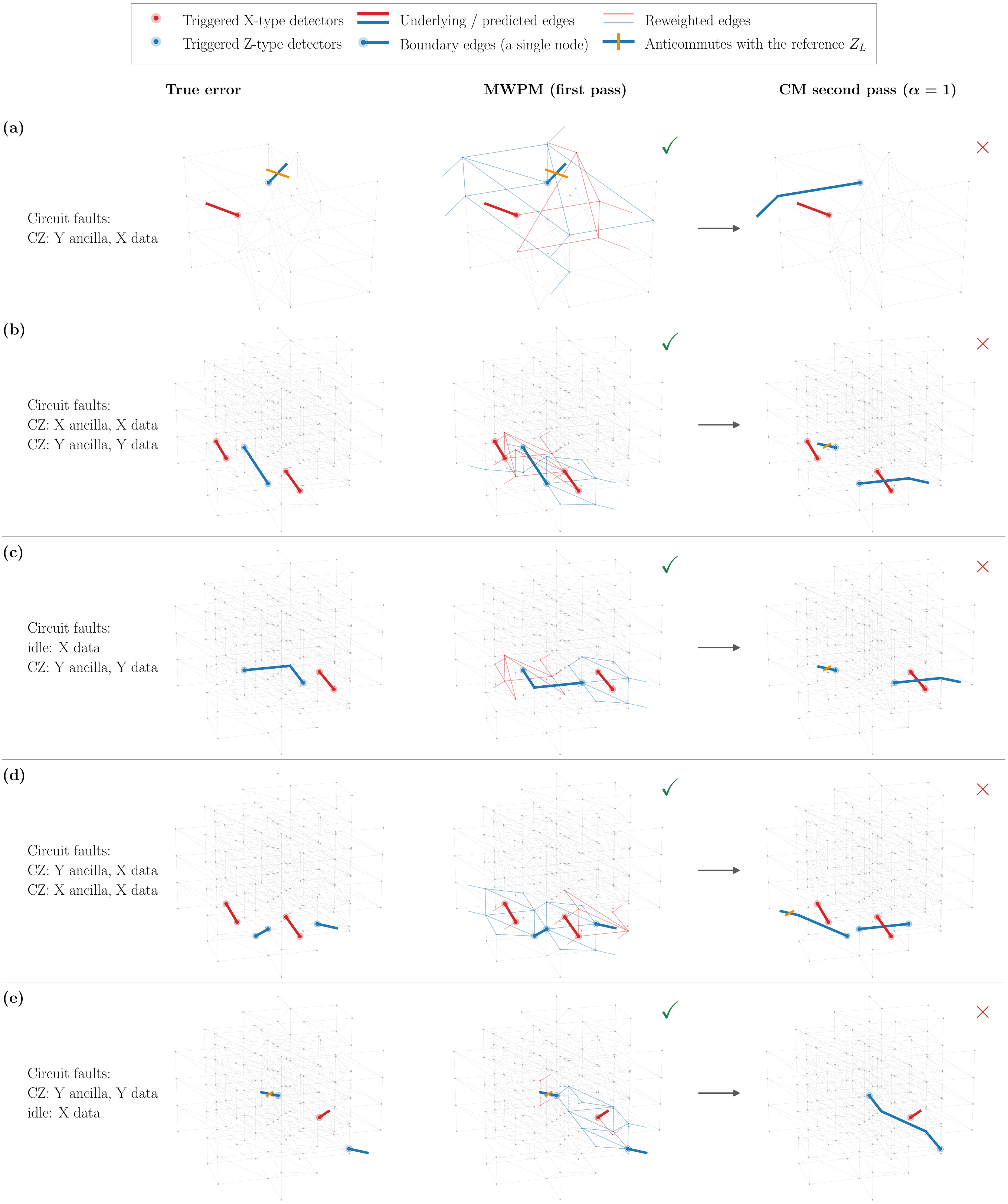}
    \caption{\textbf{Examples of size-$\frac{d-1}{2}$ CM failure modes shown in the 3D decoding graph}. The examples are at $d=3$ (a) and $d=5$ (b)--(e). 
    In all the cases shown, the failure involves a cross-basis $\textrm{CZ}$-gate error. 
    The uncorrelated (first-pass) decoder decodes each syndrome correctly, while the correlated decoder introduces a logical failure.}
    \label{fig:fault_mechanisms_stack}
\end{figure*}

Figure~\ref{fig:fault_mechanisms_stack} shows representative size-$\frac{d-1}{2}$ failures of standard correlated matching. 
From the exhaustive enumeration of size-$\frac{d-1}{2}$ fault configurations in the SI1000 $Z$-memory DEM at $p=10^{-4}$ (Appendix~\ref{app:distance}), we select examples that CM decodes incorrectly and MWPM decodes correctly.
Each row shows the decoding graph three times: the true error, the MWPM first pass ($\checkmark$), and the CM second pass at $\alpha=1$ ($\times$). 
Faint edges in the first-pass panel are those discounted for the second pass, and orange ticks mark edges that anticommute with the reference $Z_L$. 
A correction is logically correct if it crosses $Z_L$ as many times as the true error, modulo two. 
We quote the total weight $\sum_e \log[(1-p_e)/p_e]$ of each correction over both subgraphs.

In (a), at $d=3$, a single cross-basis $\mathrm{CZ}$ fault triggers one detector in each subgraph, each matched to the boundary. 
The first pass selects these two boundary edges, with a total weight of $13.28$. 
The selected $X$-type edge then discounts two $Z$-type edges that form an alternative path to the boundary, from $7.70$ and $7.82$ to $2.21$ and $2.23$. 
The two selected edges also discount each other, so the first-pass correction costs $10.54$ after reweighting. 
CM replaces the $Z$-type boundary edge with the two-edge path. 
The resulting correction would cost $21.83$ under the priors, but only $9.38$ after reweighting. 
It does not cross $Z_L$, so it produces a logical error. Panels (b)--(d) show similar mechanisms at $d=5$ with two faults. 

Panel (e) shows a more extreme example: a cross-basis $\mathrm{CZ}$ fault and an idle $X$ error leave one $X$-type and two $Z$-type detections, all matched to the boundary. 
The first pass selects these three boundary edges, with a total weight of $21.52$. 
The selected $X$-type edge then discounts three $Z$-type edges that form a chain joining the two $Z$-type detections, from $8.03$, $9.21$, and $7.68$ to $3.43$, $4.14$, and $5.66$. 
Each of these discounts comes from a different $\mathrm{CZ}$ mechanism that did not occur. 
The edges of the cross-basis fault also discount each other, so the first-pass correction costs $17.88$ after reweighting. 
The chain that CM selects would cost $31.18$ under the priors, but costs only $17.66$ after reweighting, $0.22$ below the weight of the true correction, so the decoder fails.

In this case, attributing all three discounted edges to hyperedge mechanisms that contain the source edge is possible, since an odd number of such mechanisms still activates it. It would, however, require three independent error mechanisms of probability $\mathcal{O}(p)$ to occur together, an event of order $p^3$, whereas a single mechanism of order $p$ suffices to explain the source edge. 
The second pass ignores this: it assigns each discounted edge a weight of order $\mathcal{O}(1)$, as if all three were explained by the source edge alone. 
Hence, the standard reweighting protocol prices a correction of probability $\Ord(p^3)$ below the true one, which is a two-fault configuration of order $p^2$.

In all cases, conditional weights of order $O(1)$ replace prior weights that grow as $\log(1/p)$, so one source edge can make several components of an alternative correction cheap at once. 
Each over-discounted edge removes roughly the cost of one independent fault. 
A logically inequivalent correction can therefore become cheaper than the true one already for configurations of $\frac{d-1}{2}$ faults, which a distance-$d$ code should correct.

\section{Optimal \texorpdfstring{$\astar$}{A*-} fits}
\label{sm:regularization-fits}

\begin{figure*}[t!]
    \centering
    \begin{minipage}[t]{0.48\textwidth}
        \centering
        \textbf{I.}\\[2pt]
        \includegraphics[width=\linewidth]{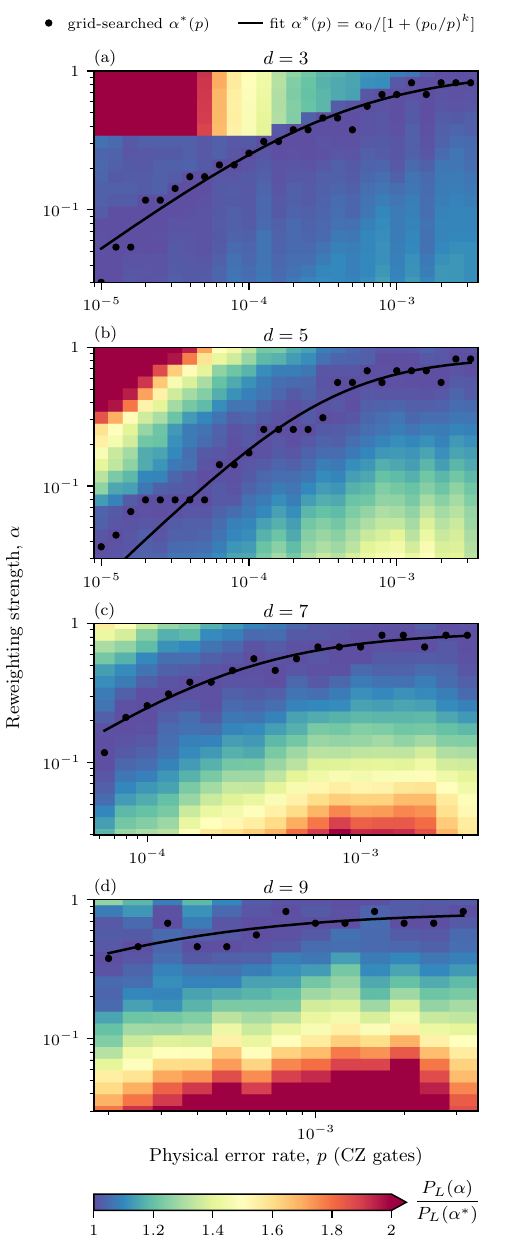} 
    \end{minipage}\hfill
    \begin{minipage}[t]{0.48\textwidth}
        \centering
        \textbf{II.}\\[2pt]
        \includegraphics[width=\linewidth]{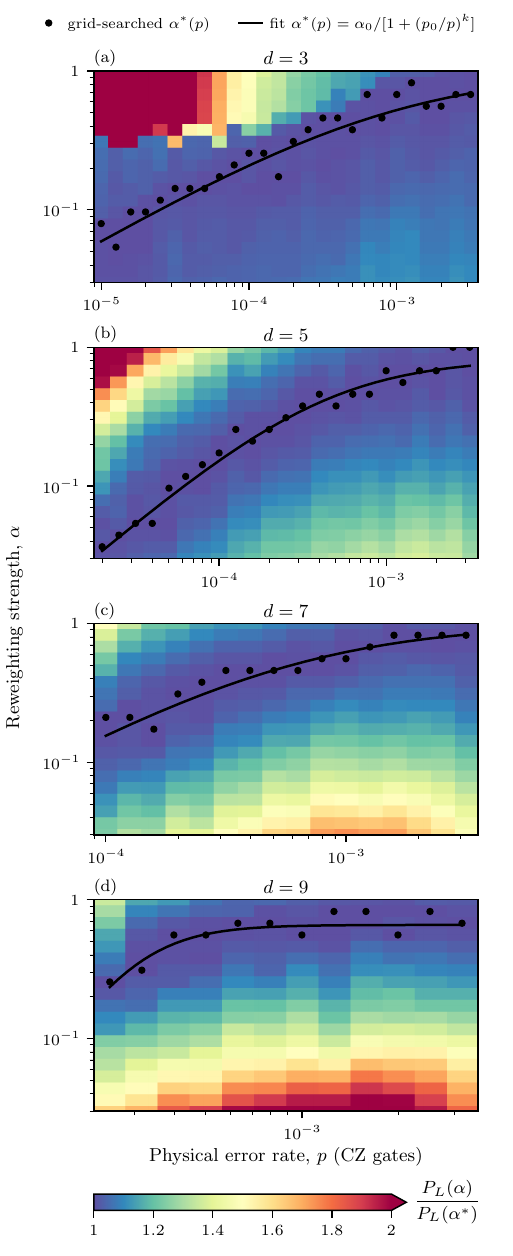}
    \end{minipage}
    \caption{\textbf{Grid search for optimal $\astar$ on simulated rotated planar surface code memory, using (I.) regularized correlated matching (RCM) and (II.) regularized correlated LCD (RC-LCD).} 
    The color shows the ratio of the logical error probability $\PL(\alpha)$ achieved with given $\alpha$ versus $\PL(\astar)$ achieved with the best candidate $\alpha$ for that error rate $p$. 
    Both MWPM and LCD show similar $\alpha$-landscapes, indicating that the optimal regularization strength is largely a property of the detector error model rather than being decoder-dependent.}
    \label{fig:alpha_landscapes}
\end{figure*}

Figure~\ref{fig:alpha_landscapes} shows the ratio of logical error rates $\frac{\PL(\alpha)}{\PL(\astar)}$ for regularized correlated matching and regularized correlated LCD with a candidate $\alpha$ versus the optimal $\astar$ (the minimum found in the sweep).
The error rates are analyzed for code distances $d=\{3, 5, 7, 9\}$. 
We see a broad valley for the optimal regularization strength $\astar$ at $d>3$, with the observed minimum following the ansatz of Eq.~\eqref{eq:alphastar}.
At $d=3$ and $p\lesssim10^{-4}$, the logical error rate rises sharply for $\alpha>0.4$.
A similar pattern appears at all sampled distances for both LCD and MWPM, with the valley narrowing as $d$ increases.
The variable landscape shows that a $p$-dependent $\alpha$ is needed---using a constant $\alpha$ across the sampled error rates induces more logical errors.
The landscape is asymmetric around $\astar$: setting $\alpha$ too high is more detrimental than setting it too low.

\begin{table}[h]\centering
\caption{Fit parameters for the optimal reweighting strength model:}
\par\smallskip
$\displaystyle \astar(p)=\frac{\alpha_0}{1+(p_0/p)^k}$
\par\smallskip
\vspace{2pt}
\begin{tabular}{llcccr}
\toprule
decoder & $d$ & $\alpha_0$ & $p_0$ & $k$ & $R^2$ \\
\midrule
CM & 3 & $1.0 \pm 0.1$ & $(4 \pm 1)\times10^{-4}$ & $0.78 \pm 0.09$ & 0.968 \\
CM & 5 & $0.85 \pm 0.08$ & $(3.5 \pm 0.9)\times10^{-4}$ & $1.0 \pm 0.2$ & 0.947 \\
CM & 7 & $0.87 \pm 0.06$ & $(2.4 \pm 0.4)\times10^{-4}$ & $1.1 \pm 0.2$ & 0.951 \\
CM & 9 & $0.8 \pm 0.2$ & $(2.0 \pm 0.8)\times10^{-4}$ & $1.0 \pm 0.8$ & 0.600 \\
\addlinespace
C-LCD & 3 & $0.8 \pm 0.1$ & $(3 \pm 2)\times10^{-4}$ & $0.8 \pm 0.2$ & 0.915 \\
C-LCD & 5 & $1.0 \pm 0.2$ & $(6 \pm 3)\times10^{-4}$ & $1.0 \pm 0.2$ & 0.938 \\
C-LCD & 7 & $1.0 \pm 0.1$ & $(6 \pm 2)\times10^{-4}$ & $0.9 \pm 0.2$ & 0.958 \\
C-LCD & 9 & $0.70 \pm 0.03$ & $(3.1 \pm 0.2)\times10^{-4}$ & $3.1 \pm 0.9$ & 0.869 \\
\addlinespace
\bottomrule
\end{tabular}
\end{table}

\end{document}